\documentclass[twocolumn, showpacs, amsmath, amssymb, nobibnotes, aps, pre]{revtex4}
\usepackage{graphicx}
\usepackage{url}
\usepackage{float}
\usepackage{latexsym}
\usepackage{amsmath,amssymb}
\usepackage{dcolumn}
\usepackage{epsfig}

\begin{document}
\title{Opinion formation in time-varying social networks: \newline The case of the naming game}

\author{Suman Kalyan Maity}
\email{sumankalyan.maity@cse.iitkgp.ernet.in}
\affiliation{Department of Computer Science and Engineering,\\ Indian Institute of Technology, Kharagpur, India -- 
721302}
\author{T. Venkat Manoj}
\email{manoj.venkat92@gmail.com}
\affiliation{Department of Computer Science and Engineering,\\ Indian Institute of Technology, Kharagpur, India -- 
721302}
\author{Animesh Mukherjee}
\email{animeshm@cse.iitkgp.ernet.in}
\affiliation{Department of Computer Science and Engineering,\\ Indian Institute of Technology, Kharagpur, India -- 
721302}

\begin{abstract}\label{abstract}
We study the dynamics of the naming game as an opinion formation model on time-varying social networks. This agent-based model captures the essential features of the agreement dynamics by means of a memory-based
negotiation process. Our study focuses on the impact of time-varying properties of the social network of the agents on the naming game dynamics. In particular, we perform a computational exploration of this model using
simulations on top of real networks. We investigate the outcomes of the dynamics on two different types of
 time-varying data - (i) the networks vary on a day-to-day basis and (ii) the networks vary within very short intervals of time (20 seconds). In the first case, we find that networks with
strong community structure hinder the system from reaching global agreement; the evolution of the naming game in these networks maintains clusters of coexisting opinions indefinitely leading to metastability.
In the second case, we investigate the evolution of the naming game in perfect synchronization with the time evolution of the underlying social network shedding new light on the traditional emergent properties of the game
that differ largely from what has been reported in the existing literature.  
\end{abstract}
\pacs{89.75.-k, 05.65.+b, 89.65.Ef}
\maketitle

\section{Introduction}
Social networks are inherently dynamic. Social interactions and human activities are intermittent, the neighborhood of individuals moving over a geographic space evolves over
time, links appear and disappear in the World-Wide-Web. The essence of social network lies in its time-varying nature. Links may exist for a certain time period and may be recurrent.
In summary, as time progresses, the societal structure keeps changing. Similarly, with the evolution of time, social conventions, shared cultural and linguistic patterns reshape themselves.
Opinions spread, some get trapped into communities, some cross the barrier of local groups/communities and become accepted globally among different communities and some die competing with others. Most of these 
social phenomena can be modeled and analyzed in a time-varying framework. Almost all previous work is limited to the analysis of the naming game dynamics on static networks
~\cite{baronchelli:06a,topo,qu,dall,liu,chaos,luca,luc,yan}. Therefore, in this paper, we focus on the competing opinion formation over time-varying real-world social networks. One way of viewing at time-varying networks is as a series of static graphs accumulated
over a fixed time interval; however these kind of networks do not always perfectly capture the temporal ordering of the links appearing in the system which may sometimes lead to over/under-estimation of network topologies.
 Thus, we plan to investigate the opinion formation process on both the accumulated static graphs as well as on its detailed time-resolved counterpart.

In this paper, we focus on the basic Naming Game model (NG)~\cite{baronchelli:06a} to study how opinions 
spread with time and how societies move towards consensus in the adoption of a single opinion through negotiation or agree upon multiple opinions due to non-uniform interaction pattern among different communities. The evolution of the system
in this model takes place through the usual local pairwise interactions among artificial agents that necessarily capture the generic and essential features of an agreement process.
This model was expressly conceived to explore the role of self-organization in the evolution of languages~\cite{steels96aSelf,steels96selfOrganizing} and has acquired a paradigmatic role in semiotic
dynamics that studies evolution of languages through invention of new words, grammatical constructions and more specifically, through adoption of new meaning for different words. NG finds wide applications in various fields
ranging from artificial sensor network as a leader election model~\cite{baronchelli:11} to the social media as an opinion formation model.

\begin{figure}[h]
\begin{center}
\includegraphics*[width=1\columnwidth,angle=0]{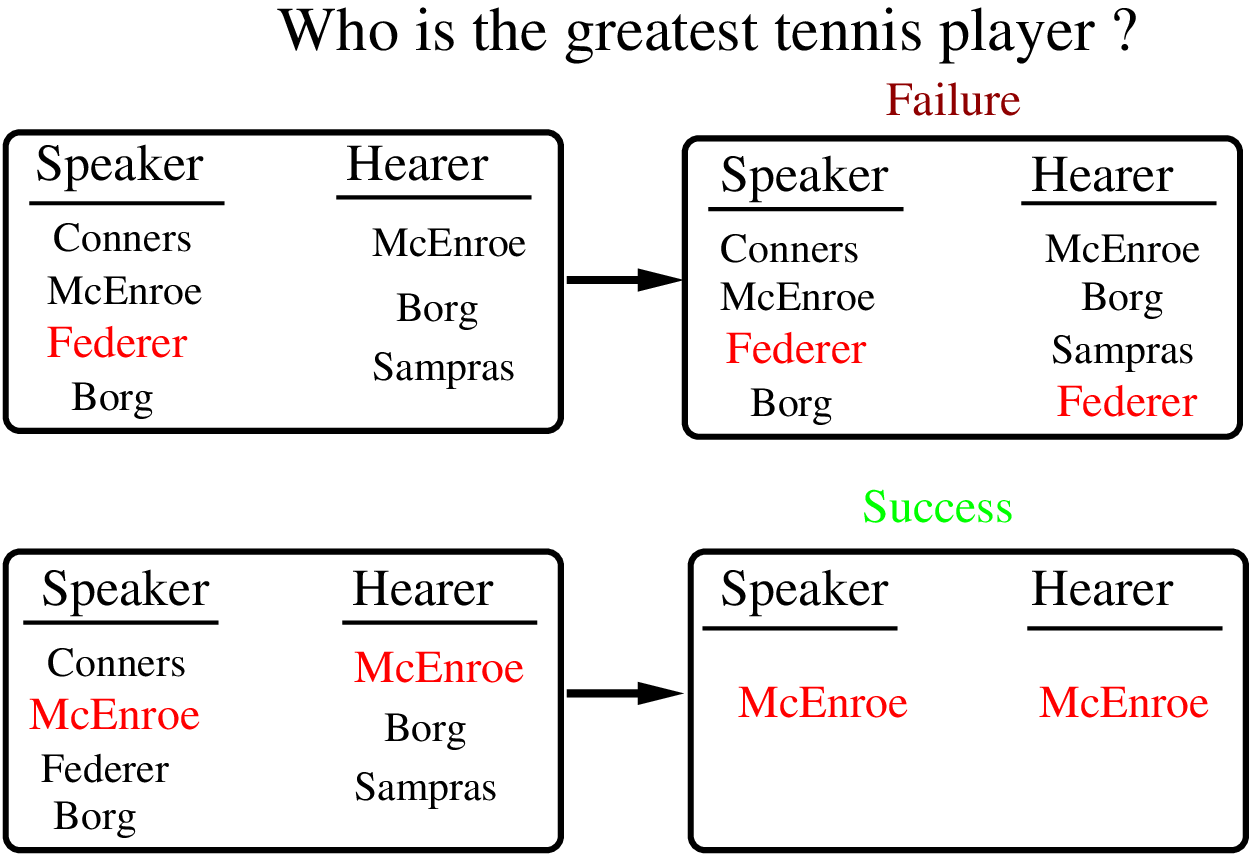}
\caption{\label{fig1}(Color online) Agent's interaction rules in basic NG. Suppose there is a topic on which a discussion is going on, say ``Who is the greatest tennis player?''. (Top) The speaker chosen at random, opines for ``Federer'' (also chosen randomly from his inventory of opinions). Now, the hearer (again chosen at random) does not have this opinion in her inventory, and therefore she adds the opinion ``Federer'' in her inventory and the interaction is a failure
. (Bottom) The speaker opines for ``McEnroe'' and in this case the opinion is present in the hearer's inventory. Therefore, they delete all other opinions except ``McEnroe''. The interaction this time is a success.}
\end{center}
\end{figure}

The minimal naming game consists of a population of $N$ agents observing a single object in the environment (may be a discussion on a particular topic) and opining for that by means of communication with one another through pairwise interactions, in order to reach a global agreement.
The agents have at their disposal an internal inventory, in which they can store an unlimited number of different words or opinions. At the beginning, all the individuals have empty inventories. At each time step, the dynamics consists of a
pairwise interaction between randomly chosen individuals. The chosen individuals can take part in the interaction as a ``speaker'' or as a ``hearer.'' The speaker voices to the hearer a possible
opinion for the object under consideration; if the speaker does not have one, i.e., ͑his inventory is empty͒, he invents an opinion͔. In case where he already has many opinions ͑stored in his inventory͒, he chooses one of them randomly. The hearer's move is deterministic: if she possesses the opinion pronounced by the speaker,
the interaction is a ``success'', and in this case both speaker and hearer retain that opinion as the right one, removing all other competing opinions/words from their inventories; otherwise, the new opinion is included in the inventory of the hearer, without any cancellation of opinions in which case the interaction is termed as a ``failure'' (see fig~\ref{fig1}).
The game is played on a fully connected network, i.e., every agent can, in principle, communicate with every other agents, and makes the following assumption. It is assumed that there can be potentially huge number of opinions for a particular topic so that the probability that two players will ever invent the same opinion at two different times is practically negligible 
and that the environment consists of a single topic of discussion.

Although the system reaches a global consensus through the invention and decay of opinions, it is interesting to note the important differences from other opinion formation models. 
In Axelrod’s model~\cite{Axelrod_1997}, each agent is endowed with a vector of opinions, and can interact with other agents only if their opinions are already close enough; in Sznajd’s model~\cite{sznajdweron}
and in the Voter model~\cite{Krapivsky92}, the opinion can take only two discrete values, and an agent takes deterministically the opinion of one of its neighbors. Further in~\cite{Deffuant01},
opinion is modeled as a unique variable and the evolution of two interacting agents is deterministic. In the naming game model on the other hand, each agent can potentially have an unlimited number of possible discrete states
 (or opinions) at the same time, accumulating in his/her memory different possible opinions; the agents are able to ``wait'' before reaching a decision. 
Moreover, each dynamical step can be seen as a negotiation between a speaker and a hearer, with a certain degree of stochasticity.

Most previous studies of the NG model in semiotic/opinion dynamics have focused on populations of agents in which all pairwise interactions are allowed, i.e., the agents are placed on the vertices of a fully connected graph~\cite{baronchelli:06a,baronchelli08}. Apart from this mean-field case, the model has also been studied  
 on regular lattices~\cite{topo,qu}; small world networks~\cite{qu,dall,liu,chaos}; random geometric graphs~\cite{qu,lu,hao}; static~\cite{luca,luc,yan}, dynamic and adaptive~\cite{who} and empirical~\cite{con} complex networks.

In this paper, we consider the NG dynamics on two different types of time-varying data; one varying on a day-to-day basis while another varying over very short intervals of  time (20 seconds). In the first case, we observe that networks with strong community structures delay
the convergence due to co-existence of competing and long-lasting clusters of opinions. In the second case, the games are played in perfect synchronization with the time-evolution of the network. In this case, we observe that the global observables are markedly different from the case where
the games are played on the static (and composite) version of the same network as well as from the traditional results reported in the literature.

The rest of the paper is organized as follows. In Section 2, we describe the datasets on which we investigate
the naming game dynamics in a time-varying social scenario. Section 3 provides the elaborate model description. In Section 4, we present the results and provide explanations for our findings. Finally, conclusions are drawn in section 5.

%
%
 
\section{Datasets}
For the purpose of the investigation of the NG dynamics on time-varying networks, we consider two specific real-world face-to-face contact datasets and present our results on each of them. Both the datasets are obtained from SocioPatterns Collaboration 
(http://www.sociopatterns.org/datasets/). The data collection infrastructure uses active RFID devices embedded in conference badges to detect and store face-to-face proximity relations of persons wearing the badges. These devices can detect face-to-face proximity (1-1.5 meter) of individuals wearing the badge with a temporal resolution of 20 seconds. 
A detailed description of the datasets on which we conduct our experiments are as follows:

\textbf{SCIENCE GALLERY}: The dataset comprises of face-to-face interaction data of visitors of the Science Gallery in Dublin, Ireland during the 
spring of 2009 at the event of art-science exhibition ``INFECTIOUS: STAY AWAY''~\cite{Isella2011166}.

a) \textit{Cumulative data per day} ($SG_{DAY}$): This comprises of 69 weighted networks of the visitors of the Science Gallery each representing a cumulative static aggregation network of a single day. Each of these daily interaction networks are obtained by aggregating the contact sequences for a period of 24 hours.
The nodes in these networks represent visitors of the Science Gallery while an edge is drawn between two nodes if at least one contact was detected between those nodes during the interval of aggregation. Thus, every edge is weighted by the number
 of 20 seconds intervals during which the corresponding individuals have been in close-range proximity. In other words, if $i$ and $j$ remain in contact for $m$ number of 20s interval then the weight $w_{ij} = m$.

b) \textit{Time resolved data} ($SG_{SECS}$): This dataset consists of time-varying versions of the networks for each of the 69 days. On each day, the evolution of the face-to-face interactions of the agents is captured by varying snapshots of the interaction network obtained after every 20 second time interval.

\textbf{HYPERTEXT, 2009} ($HT_{SECS}$): The face-to-face interaction data of the conference attendees of ACM Hypertext 2009 held in Institute for Scientific Interchange Foundation in Turin, Italy,
 from June $29^{th}$ to July $1^{st}$, 2009, where the SocioPatterns project deployed the Live Social Semantics application. The dataset contains the dynamical network of face-to-face proximity of 113 conference attendees over about 2.5 days.

\section{The model description}
The basic NG Model can be summarized as follows. At each time step ($t$ = 1, 2, ..) two agents are randomly selected to interact: one of them plays the role of speaker, the other one that of hearer. The interactions obey the following rules
\begin{itemize}
 \item The speaker voices an opinion from his list of opinions to the hearer. (If the speaker has more than one opinion on his list, he randomly chooses one; if he has none, he invents a brand new opinion~\footnote{ For implemetation purposes this refers to maintaining a word identification number which is incremented by one for every new invention.})
\item If the hearer has this opinion, the communication is termed ``successful'', and both players delete all other opinions, i.e., collapse their list of opinions to this one opinion. Therefore, they meet a local agreement.
\item If the hearer does not have the opinion transmitted by the speaker (termed ``unsuccessful'' communication), she adds the opinion to her list of opinions without any deletion.
\end{itemize}
Note that in this model any agent is free to interact with any other agent, i.e., the underlying social structure is assumed to be fully connected. For the purpose of our analysis however, we assume that the agents are embedded on 
realistic social networks (i.e., SG and HT) that are continuously varying over time. In this case, although the basic rules of the game remain exactly same, the only issue is to devise a strategy for the speaker-hearer selection.
 We consider two variants of this selection, the first one being suitable for the $SG_{DAY}$ dataset and the second one for the $SG_{SECS}$ and $HT_{SECS}$ dataset.

\textbf{Strategy I (weighted version):} Here we randomly select a speaker and preferentially choose a hearer among his neighbors.
 Our intention is to simulate an important criterion: we talk most preferably to those with whom we had already met before. This is implemented as follows:
\begin{itemize}
 \item The speaker $i$ is selected randomly.
\item The hearer $j$ is chosen from among the neighbors of $i$ such that it is more likely that $i$ shares a high-weighted link with $j$, i.e., the choice is according to the probability 
                                                                               \begin{equation*}
                                                                              p_{ij} = \frac{w_{ij}}{\sum_{j=1}^{k}w_{ij}}
                                                                               \end{equation*}
where $w_{ij}$ (see section II for definition) is the edge weight between the pair $i$ and $j$ while $k$ is the number of neighbors of $i$.
\end{itemize}
\textbf{Strategy II (unweighted version):} This variant is quite straight-forward. We choose a random speaker and a random hearer among his neighbors to impart equal importance to each pair of connections. 

The main quantities of interest which describe the emergent properties of the system are
\begin{itemize}
 \item the total number $N_w(t)$ of words/opinions in the system at the time $t$ (i.e., the total size of the memory);
\item the number of different words/opinions $N_d(t)$ in the system at the time $t$;
\item the average success rate $S(t)$, i.e., the probability, computed averaging over many simulation runs, that the chosen agent gets involved in a successful interaction at a given time $t$.
\end{itemize}

From a global perspective, the quantities which are of interest are the time to reach the global consensus ($t_{conv}$), the maximum memory required by the agents during the process ($N_w^{max}$) and 
the time required to reach this memory peak ($t_{max}$).
 
\section{Results and discussions}
In this section, we present the results of the analysis of the NG dynamics on the $SG_{DAY}$, $SG_{SECS}$ and the $HT_{SECS}$ datasets.
\subsection{Analysis on day-wise $SG_{DAY}$ dataset}
We have studied the opinion formation process on the 69 days of close interactions among the visitors for the $SG_{DAY}$ dataset. We play the naming game on each of the composite daily networks of 69 days seperately following Strategy I until we reach consensus.
 Note that we are still playing NG on static and composite versions of the networks obtained at the end of each of the 69 days seperately. The primary focus of this study is to find the day-to-day differences in the behavior of the global quantities obtained by playing the NG until consensus is reached on the top of each of the 69 instances individually.
 These composite instances can be thought of as 69 different societal structures (with a certain amount of overlap in the nodes between any two consecutive days) and the objective here is to investigate how the day-to-day variability of this structure affect the opinion formation in terms of the global quantities of interest.
The sizes of the giant component in these networks are not always $O(N)$. Therefore, for such disconnected networks, the NG will never reach consensus on only one opinion because the network is inherently broken into several connected components with no links among them to propel opinion shift. In particular, $N_d(t)$ never converges to 1. Therefore, in such cases we redefine $t_{conv}$ as the time to reach the following state: 
$N_{w}(t) = N $ and $N_d(t) = c$ where $c$ is the number of disconnected components (signifying one opinion per component as the convergence criteria). In the rest of this section, we shall analyze how the opinion dynamics gets affected as the underlying network structure varies on a day-to-day basis.
\subsubsection{Analysis of $N_w^{max}$} 
\begin{figure}[h]
\begin{center}
\includegraphics*[width=1\columnwidth, angle=0]{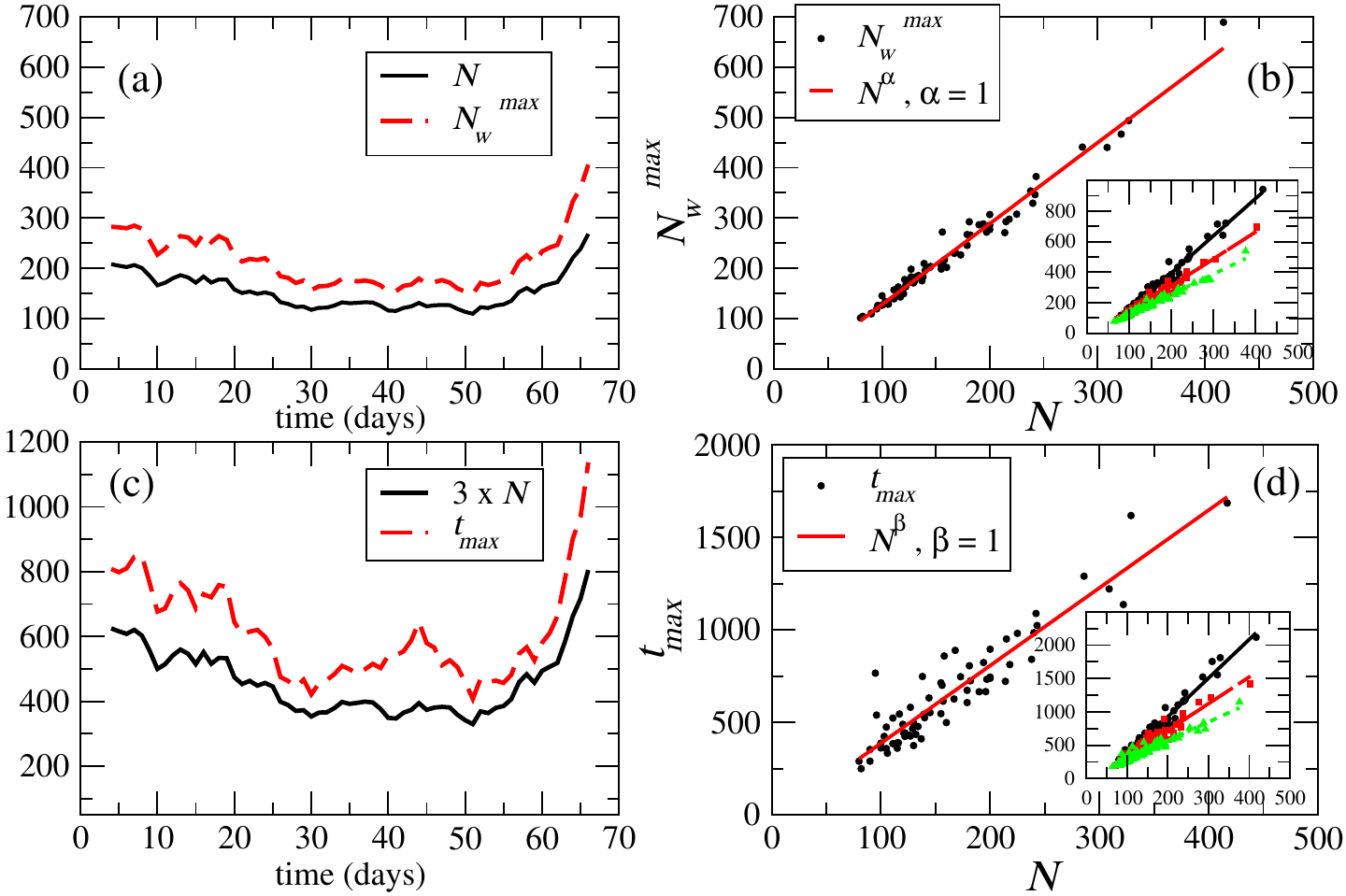}
\caption{\label{fig2}(Color online) Scaling relation of $N_w^{max}$ and $t_{max}$ with population size $N$. (a) Temporal behavior of $N_w^{max}$ and the population size 
$N$ for each of the 69 instances. Data are smoothed by taking sliding window moving average with window size of 7. (b) Scaling of $N_w^{max}$ with $N$ which is linearly fitted. The inset shows the scaling 
of $N_w^{max}$ in networks where the edges with weights below values 1, 2 and 5 respectively are filtered out. All these curves are linearly fitted. (c) Variation of $t_{max}$ and population size $N$ with time. Data are smoothed taking sliding window moving average with window size of 7.
(d) Scaling of $t_{max}$ with $N$ which is linearly fitted. The inset shows the scaling of $t_{max}$ in networks where the edges with weights below values 1, 2 and 5 respectively are filtered out. All these curves are linearly fitted.}
\end{center}
\end{figure}

One of the important observables for the NG dynamics is the memory peak, i.e., the maximum memory required by the system during the course of the dynamics. It is interesting to note how this memory peak $N_w^{max}$
  varies with the population size $N$ (see fig~\ref{fig2}(a) \footnote{Moving average is a standard smoothing technique for time-series data. Moving average of window size $m$ refers to a series of arithmetic means, each of $m$ successive observations
of the given data and are shown against the midpoints of the time intervals covered by the respective group. For example, to begin with, we take first $m$ values (i.e., the window); at the next stage, exclude the first and include the $(m+1)^{th}$ value and so on. We repeat this until we reach the last set of $m$ values.}) over the days. In particular, this quantity has a linear scaling with the population size $N$ (see fig~\ref{fig2}(b)). This relationship is in agreement with what has been observed 
in the literature for small-world, scale-free and random networks~\cite{luca,chaos} where $N_w^{max}$ scales as $O(N)$.

\subsubsection{Analysis of $t_{max}$} Another important global quantity of the dynamics is the time required by the population of agents to reach the memory peak (i.e., $N_w^{max}$). Fig~\ref{fig2}(c) once again shows a strong correlation of $t_{max}$ with the population size $N$. The relationship is further explored in fig~\ref{fig2}(d) where a linear scaling is observed between them.
 This observation is also in perfect agreement with the existing literature.
\subsubsection{Analysis of $t_{conv}$}
\begin{figure}[h]
\begin{center}
 \includegraphics*[scale=0.5]{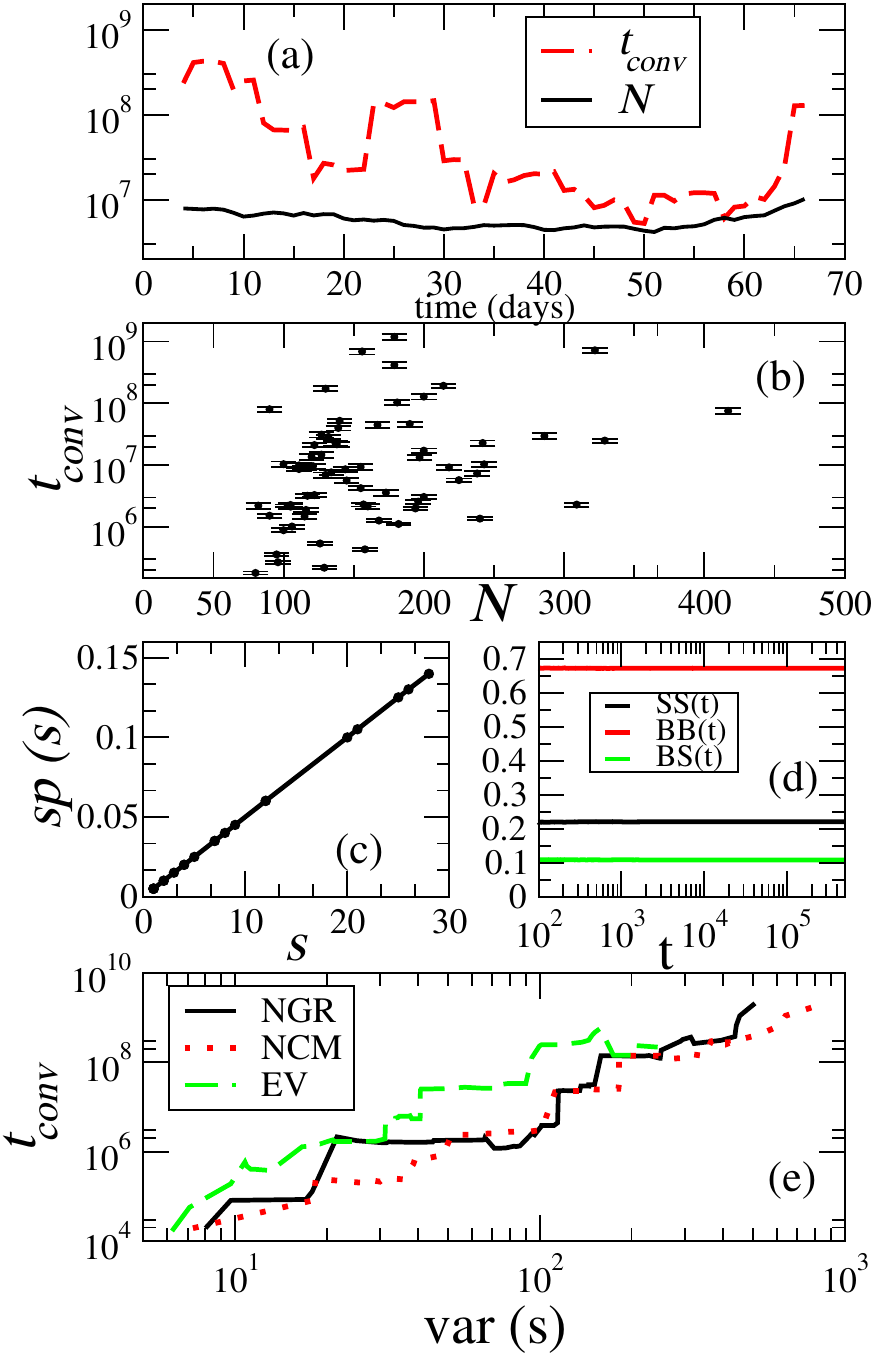}
\caption{\label{fig3}(Color online) (a) Temporal behavior of $t_{conv}$ and the population size $N$. Data are smoothed by taking sliding window moving average with window size 7. (b) Scatter plot of $t_{conv}$ versus $N$. Each datapoint is the average of 100 simulation runs. The standard error for each datapoint is plotted as errorbars. (c) Variation of $sp(s)$ vs $s$ at $t = 10^5$ (representative) for day 9 where $s$ is the community size and $sp(s)$ is the fraction of times the agents belonging to a community of size $s$
 are chosen as speakers. (d) Temporal behavior of $SS(t)$, $BB(t)$ and $BS(t)$ for day 9 data. (e) Correlation of $t_{conv}$ with variance of community sizes var(s) detected by various community detection algorithms NGR, NCM and EV. The curves are smoothed by taking sliding window moving average with window size 20.}
\end{center}
\end{figure}

The most important observable for the NG dynamics is the time to reach the global agreement $t_{conv}$. Unlike the other observables, $t_{conv}$ does not show a direct correspondence with the population size $N$ (see fig~\ref{fig3}(a) and (b)). This behavior of $t_{conv}$ is not in lines of the existing literature 
where it is usually noted that $t_{conv} \sim N^{1.4}$. Therefore the natural question that needs to be addressed is that what is (are) the property(s) of the underlying network that leads to such a non-conforming behavior of $t_{conv}$.

In fact, the answer to this question lies in the common behavior of real-world social networks. These networks typically consist of a number of communities; nodes within communities are more densely connected, while links bridging communities are sparse.
 The effect of the community structure plays a dominant role with the emergence of long-lasting multi-opinion states at the late stage of the dynamics which has also been observed in
 ~\cite{luca} and ~\cite{con}. We observe that each community reaches internal consensus fast but the weak connections between communities are not sufficient for opinions to propagate from one community to the other leading to long multi-opinion 
states which are also known as ``metastable states'' in the statistical physics literature. Formally, a metastable state is a state of the dynamics where global shifts are always possible but progressively more unlikely and the response properties 
depend on the age of the system~\cite{Mukherjee2011Aging}. Community structures are essentially authentic signatures of metastability that inhibits the dynamics leading to very slow convergence.

Presence of community structures slows down the dynamics, however, what renders the system even slower is the presence of different-sized communities. The reason for this is quite straight-forward: the agents that are part of a larger size community have a higher probability
of being chosen for a game than those belonging to a smaller size community. This is a reminiscent of the fact that the speakers are always chosen randomly which automatically increases the chances of landing in a larger size community simply because a larger bulk of the population is confined 
within this community (see fig~\ref{fig3}(c)). Therefore, even after consensus is reached in a large community, the system keeps on choosing agents from this community itself mostly resulting in ``success with no outcome''. Further, since the inter-community links are weak, and agents 
from smaller communities are rarely chosen the overall state of the system hardly changes, thereby, always keeping the agents away from the global consensus. This is also supported by the fig~\ref{fig3}(d) where we propose three metrics $SS(t)$, $BB(t)$, $BS(t)$ representing the fraction of games upto time $t$ played by a pair of agents both belonging to smaller communities,
 both belonging to bigger communities and one belonging to the bigger and other to the smaller community respectively. We see that the games are played mostly by agents both belonging to larger communities followed by agents both belonging to smaller community whereas the games which involve one agent from a large community and the other from a small community are the least in number. This phenomenon is reflected throughout the game dynamics.
Further, we report in fig~\ref{fig3}(e) the correlation of $t_{conv}$ with the variance of the community sizes. The basic idea is as follows: if a
 network gets decomposed into {\em m} communities each of size $s_1, s_2, . . ., s_m$ then we calculate the statistical variance of this size distribution and plot it against $t_{conv}$. For the purpose of community analysis, we use three standard algorithms - Newman and Girvan (NGR)~\cite{NewGir04},
 Newman, Clauset and Moore (NCM)~\cite{Clauset04} and community detection by eigen vector (EV)~\cite{Newman_2006} and in each case we observe that $t_{conv}$ has a strong positive correlation with the variance of the community sizes (see fig~\ref{fig3} (e)). 
\subsubsection{Effect of edge weights}
\begin{figure}[h]
\begin{center}
\includegraphics*[width=1\columnwidth]{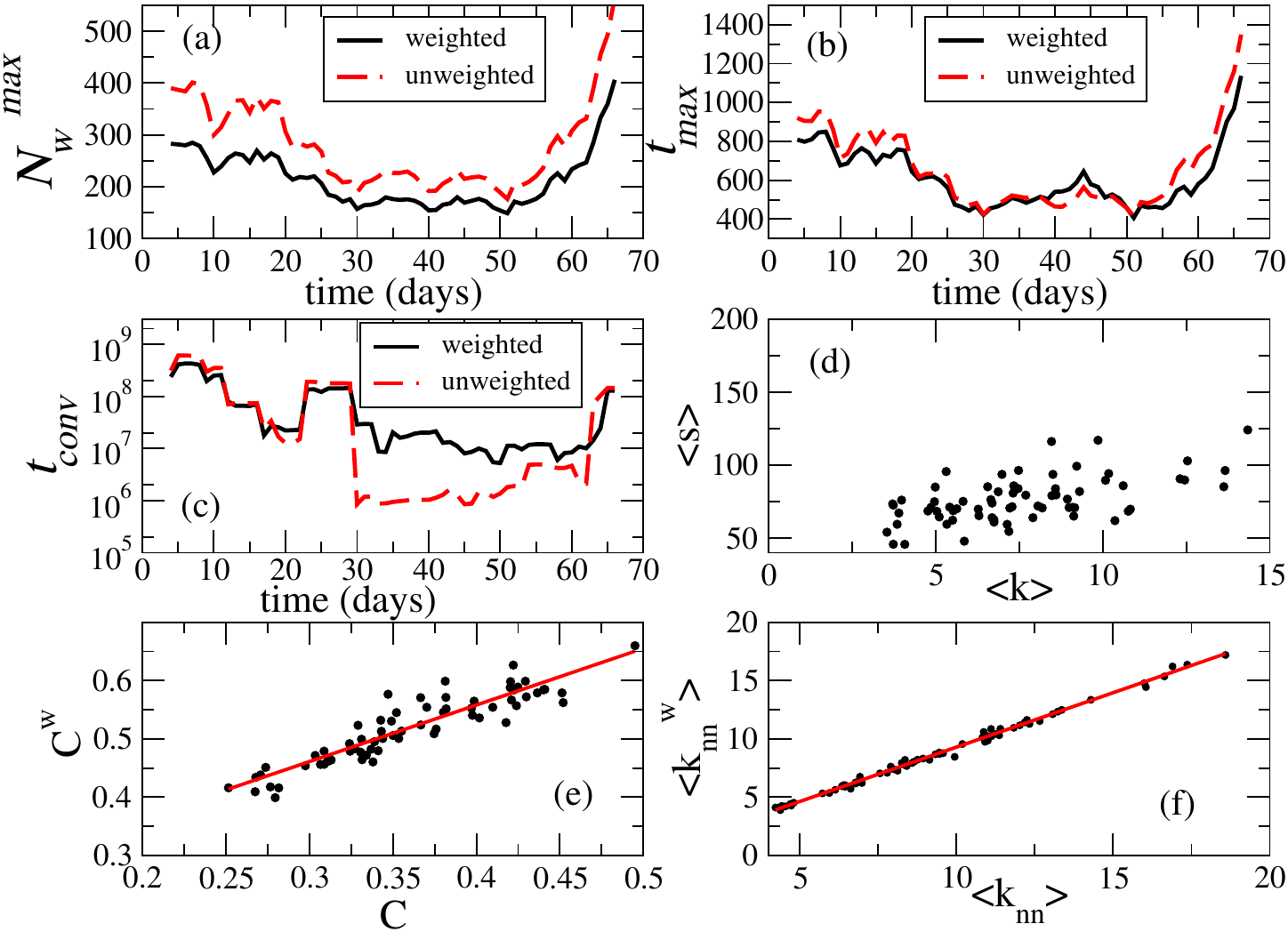}
\caption{\label{fig4}(Color online) Effect of edge weights on the dynamics. Temporal evolution of (a) $N_w^{max}$, (b) $t_{max}$ and (c) $t_{conv}$ for NG on the weighted and the unweighted network respectively. Data are smoothed by taking sliding window moving average with window size 7 .
 (d) Variation of $\langle k \rangle $ with $\langle s \rangle $. 
 (e) Variation of $C^w$ with $C$. The curve is linearly fitted ($R^2 = 0.85$). (f) Variation of $\langle k_{nn}^w \rangle$ with $\langle k_{nn} \rangle$. The curve is linearly fitted ($R^2 = 1$). }
\end{center}
\end{figure}
We observe that the naming game dynamics exhibits very similar behavior irrespective of the fact whether the underlying network is weighted or unweighted (see fig~\ref{fig4}(a), (b) and (c)). We conjecture that this is because various topological properties of the weighted and unweighted
versions of the same network are correlated to a significant extent. For instance, there is a strong correlation between the average degree $\langle k \rangle$ (unweighted network) and the average strength $\langle s \rangle$ (weighted network) (fig~\ref{fig4}(d)). The weighted clustering coefficient $C^w$ is almost perfectly correlated to the unweighted clustering coefficient $C$ (see fig~\ref{fig4}(e)).
 Further, the weighted average nearest neighbor degree $\langle k_{nn}^w \rangle$ and the unweighted average nearest neighbor degree $\langle k_{nn} \rangle$ are perfectly correlated with linear dependency in between them (see fig~\ref{fig4}(f)). All these observations can be possibly attributed to the fact that a huge majority of the links in the weighted networks
have very low weights \footnote{The average of the mean of weight distributions for all these 69 networks is 11 (i.e, majority are low-weight edges) whereas the average standard deviation of these weight distributions is 20.5 which is non-negligible indicating the presence of atleast a few high weight edges.} (on an average 67 \% of the edges have weights below 5) which make them roughly equivalent in structure to their unweighted counterparts. However, a deeper analysis on this issue remains a part of our future scope.

\subsubsection{Examples of individual instances}  
\begin{figure}[h]
\begin{center}
 \includegraphics*[width=1\columnwidth,angle=0]{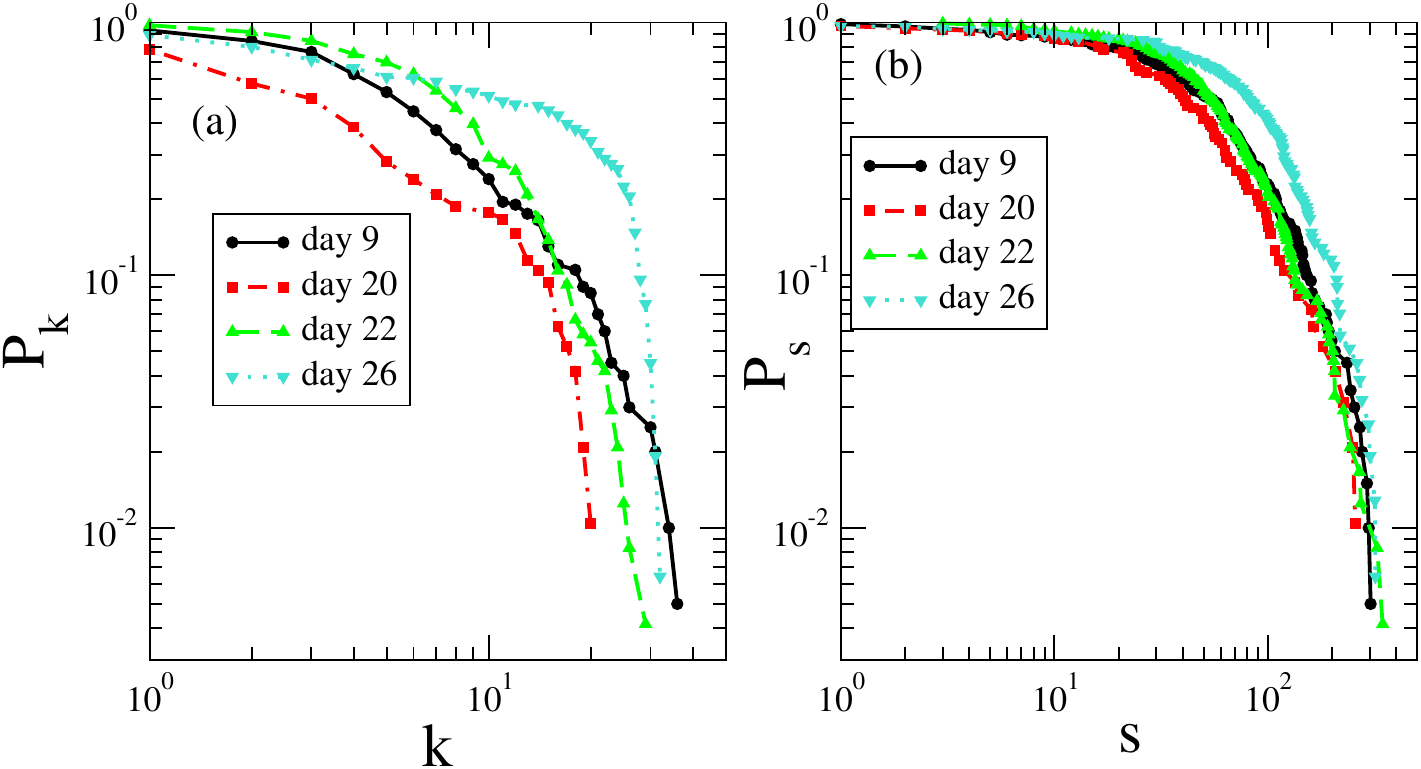}
 \caption{\label{fig4.9}(Color online) (a) Cumulative Degree distribution (greater than type) (b) Cumulative strength distribution (greater than type) for four representative networks.}
\end{center}
\end{figure}

In this subsection, we dig deeper into the individual snapshots to have a more clear understanding of the on-going dynamics. From the 69 instances,
 we present four representative cases that roughly capture all the different characteristics found across the instances. The degree and the strength (i.e., weighted degree) distributions of these networks are shown in fig~\ref{fig4.9}. Note that the degree and the strength
distributions exhibit very similar behavior for all the four networks - this is in correspondence with the observations made in the previous subsection. Both these distributions roughly indicate a small initial power-law regime followed by a cut-off. Although these statistical distributions look fairly similar, 
the outcome of the naming game dynamics as we shall see, is largely different across these networks.

Two among these, consist of disconnected components while the other two are single connected components. Further, two of them (one connected and the other disconnected) show fast convergence while the other two (again one connected and the other disconnected) show slow convergence triggered by the presence of community structures leading to metastability.  
Here we propose two metrics to capture the two distinct behaviors of the convergence time. The first one is the average number of unique words per community which is denoted by $U(t)$ and defined as follows:
\begin{equation*}
U(t) = \frac{\sum_{i=1}^{C}|A_i|}{C}  
\end{equation*} where $C$ is the number of communities and $A_i$ is the list of unique words in community $i$.

The second metric we propose is the average overlap of unique words across communities which is denoted by $O_c(t)$ and defined as follows:
\begin{equation*}
O_c(t) =\frac{2}{C(C-1)} \sum_{i>j} \frac{2(|A_i \bigcap A_j|)}{\sqrt{2(|A_i|^2+|A_j|^2)}}
\end{equation*}
\begin{figure}[h]
\begin{center}
 \includegraphics*[width=1\columnwidth,angle=0]{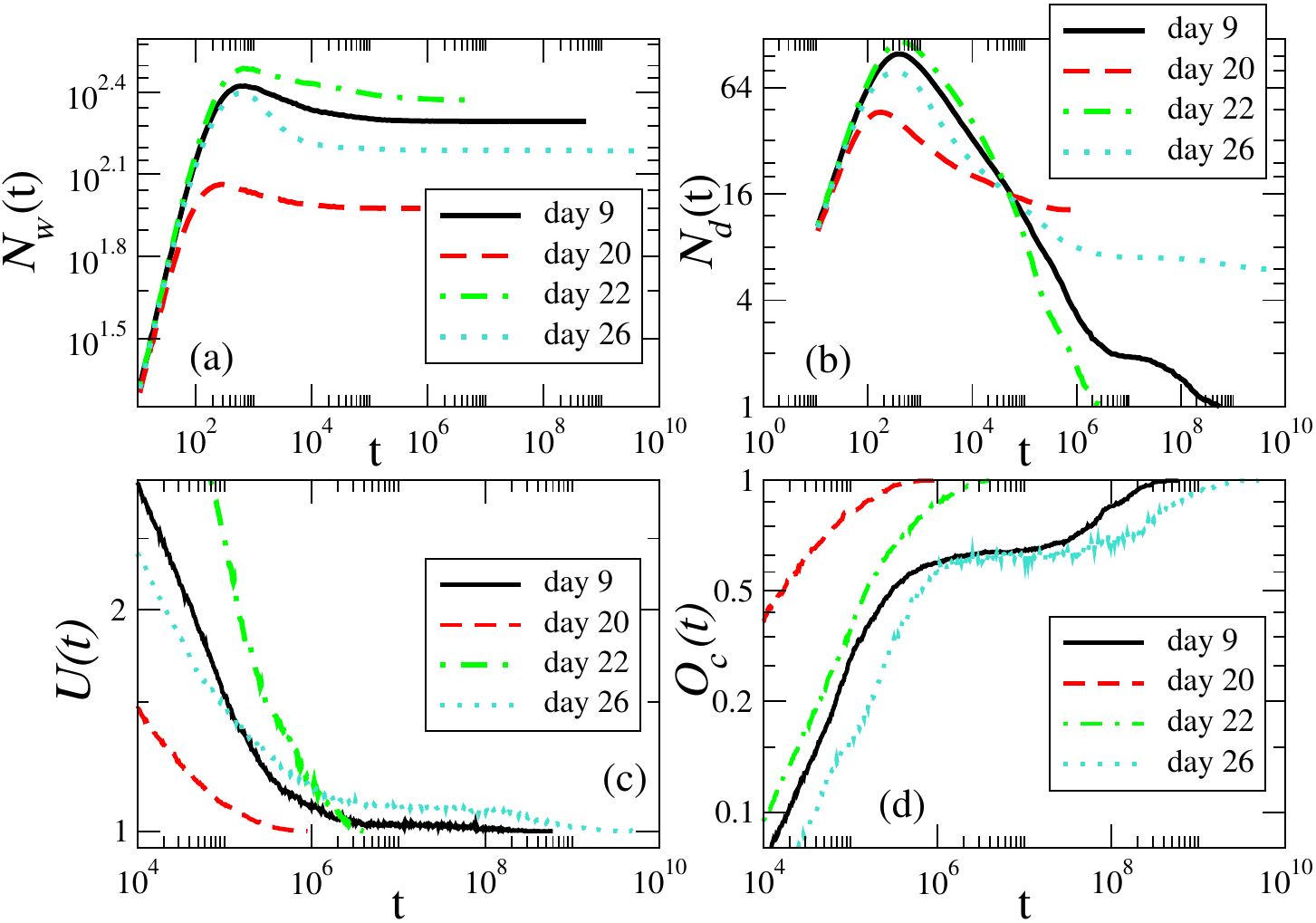}
\caption{\label{fig5}(Color online) Comparison of the evolution of the (a) total number of words $N_w(t)$ and (b) number of different words $N_d(t)$ with time on four representative networks. (c)
Average number of unique words per community $U(t)$ evolving over time. (d) Temporal evolution of average overlap of words across communities $O_c(t)$. Each point in the above curves represents the average
value obtained over 100 realizations of NG on top of each of the four networks.
}
\end{center}
\end{figure}
We consider the daily networks of $9^{th}$, $20^{th}$, $22^{nd}$ and $26^{th}$ day. The $9^{th}$ and $22^{nd}$ day network structures consist of a single connected component with 200 and 240 nodes respectively while
the $20^{th}$ and $26^{th}$ day network consist of multiple disconnected components with 96 and 156 nodes respectively. The evolution of $N_w(t)$ shows a steady growth signifying inventions of new opinions coupled with a series of failure interactions 
until the maxima is reached (see fig~\ref{fig5}(a)). From this point onward, the reorganization phase commences and the players encounter mostly successful interaction resulting in the drop of $N_w(t)$ (fig~\ref{fig5}(a)).
While for the $20^{th}$ (disconnected network) and $22^{nd}$ (connected network) day consensus is reached fast, for the $9^{th}$ (connected network) and the $26^{th}$ (disconnected network) day the system gets arrested in a long plateau indicating the 
presence of metastability and strong community structures. The growth of $N_d(t)$ also signifies similar pattern, steady rise followed by 
steady fall and a plateau (signifying a strong community structure) in case of the $9^{th}$ and $26^{th}$ day (see fig~\ref{fig5}(b)). To explore the flat plateau
region further we report $U(t)$ and $O_c(t)$ in fig~\ref{fig5}(c) and~\ref{fig5}(d) respectively. It is interesting to note that both $U(t)$ and $O_c(t)$ show a plateau in case of the $9^{th}$ and $26^{th}$ day which is a signature of the fact 
that the games played in the plateau region predominantly produce success with no deletion of opinions leading to the emergence of metastability. It should be remarked here that one can analyze these two metrices $U(t)$ and $O_c(t)$ for the NG dynamics to find signatures of community structures across different social networks.

\subsection{Analysis on the time-resolved dataset}
In this section, we consider the datasets containing dynamic face-to-face interactions. We play the naming game on these time-varying networks in complete synchronization with
the real time i.e., at each time step $t$ = 1, 2 . . . , (the elementary unit of time being second) a game is played among those agents that are alive (having degree atleast 1) at that particular instant of time in the network. In other words, at every time step (i.e., one second) one iteration of the game is played choosing a couple according to Strategy II 
(a random speaker from the active agents and a random hearer among his neighbors). When the system moves from a given network, say at time $t$, to the next one, at $t+1$, the portion of the old link structure that still prevails at $t+1$ as well as the opinion states of all the agents are retained.

\subsubsection{Results from $SG_{SECS}$ dataset}
In this section, we present the results of the naming game dynamics on the time-resolved $9^{th}$ day network from the $SG_{SECS}$ dataset. Note that the results that we present in the following are very robust and perfectly representative of the results that one would obtain in case of the time-resolved data for any other day.
\begin{figure}[h]
\begin{center}
 \includegraphics*[width=1\columnwidth,angle=0]{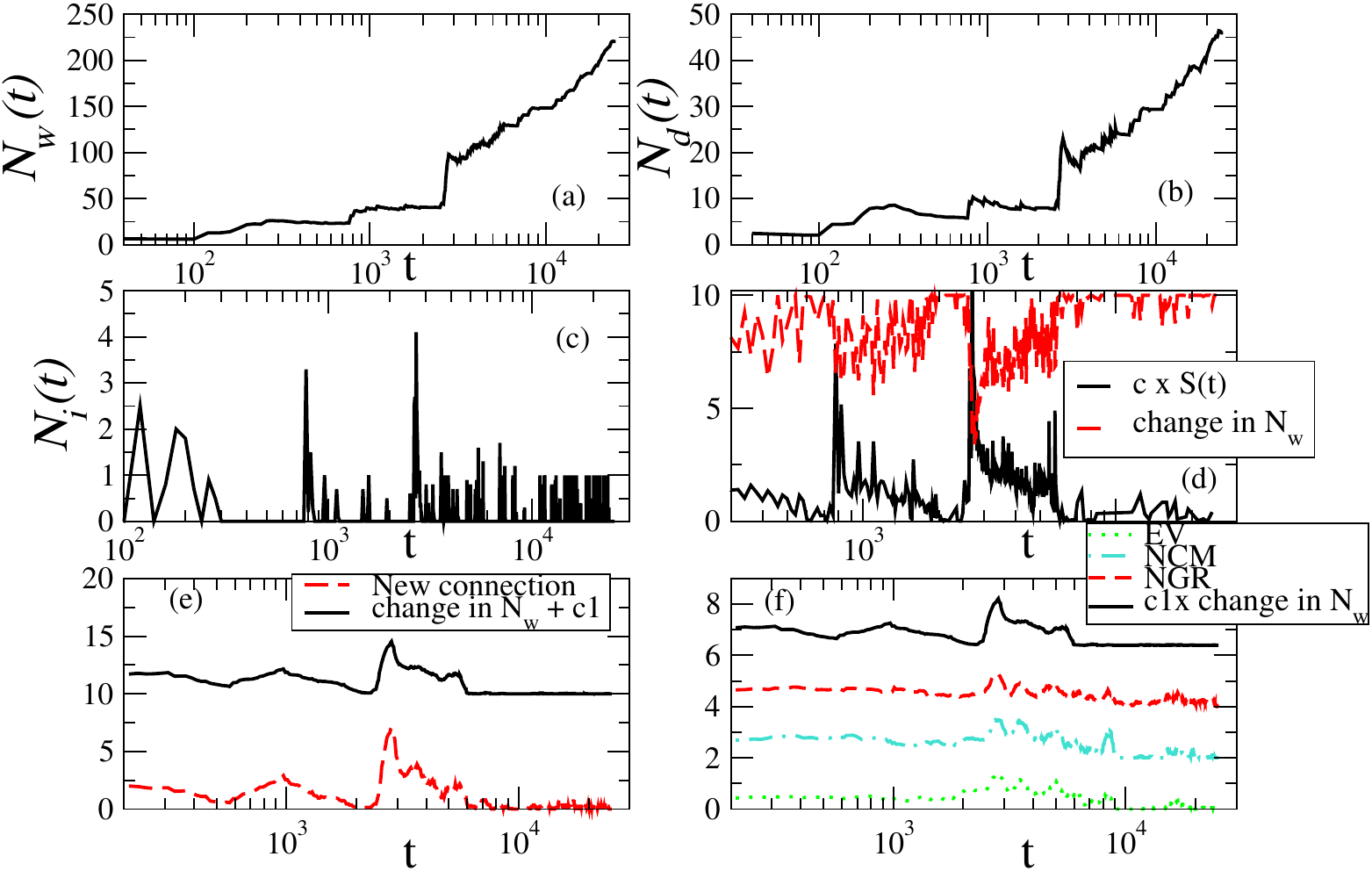}
\caption{\label{fig6}(Color online) The temporal evolution of (a) $N_w(t)$ and (b) $N_d(t)$ for day 9 ($SG_{SECS}$ dataset). The data are averaged over 100 simulation runs. 
(c) The number of inventions of opinions $N_i(t)$ over time. (d) Comparison of $\Delta N_w(t)$ with success rate $S(t)$. (e) Comparison of temporal evolution of $\Delta N_w(t)$ and number of new connections smoothed by taking sliding window moving average with window size 20.
 (f) Comparison of $\Delta N_w(t)$ with the variance of community sizes (found by NGR, NCM and EV algorithm) evolving over time (the curves are suitably scaled by some constant for the purpose of better visualization). The data are smoothed by taking sliding window moving average with window size 20.}
\end{center}
\end{figure}

The time evolution of $N_w(t)$ and $N_d(t)$ on the time-varying graph of day 9 (see fig~\ref{fig6}(a) and (b)) show a drastically different behavior from the case where these quantities are measured on the static (and composite) counterpart (see fig~\ref{fig5}(a) and (b)). The temporal graph shows a slow growth regime followed by a sharp transition, whereas the static counterpart shows steady growth regime followed by a steady fall and finally a long-lasting metastable state
(see fig~\ref{fig5}(a)). This difference in behavior is due to the fact that in the time-varying case inventions of opinions prevail throughout the dynamics (see fig~\ref{fig6}(c)) which prevents the disposal of opinions from the system and hence the memory sizes do not decrease.
Further, in fig~\ref{fig6}(d) we show how the absolute change in $N_w$ is driven by the success rate; $\Delta N_w(t)$ increases with a decrease in $S(t)$ while it decreases with an increase in $S(t)$. Fig~\ref{fig6}(e) shows the direct dependence of $\Delta N_w$ on the number of new connections.
 Another interesting property which has an impact on the dynamics is the community size. Indeed the variance of the community sizes are in correspondence with $\Delta N_w$ (see fig~\ref{fig6}(f)). For community analysis, we consider each of the snapshots at 20 seconds interval and apply NGR, NCM and EV individually on these snapshots. Further, we consider only the active nodes (having degree atleast 1) for the purpose of community detection. 

Therefore, in a nutshell, the continuous inventions, the influx of new connections (causing more failures) and the fact that the opinions 
get trapped within local neighborhoods together contribute to the steeply rising memory size over the time evolution of the dynamics.
\begin{figure}[h]
\begin{center}
\includegraphics*[scale=0.3,angle=0]{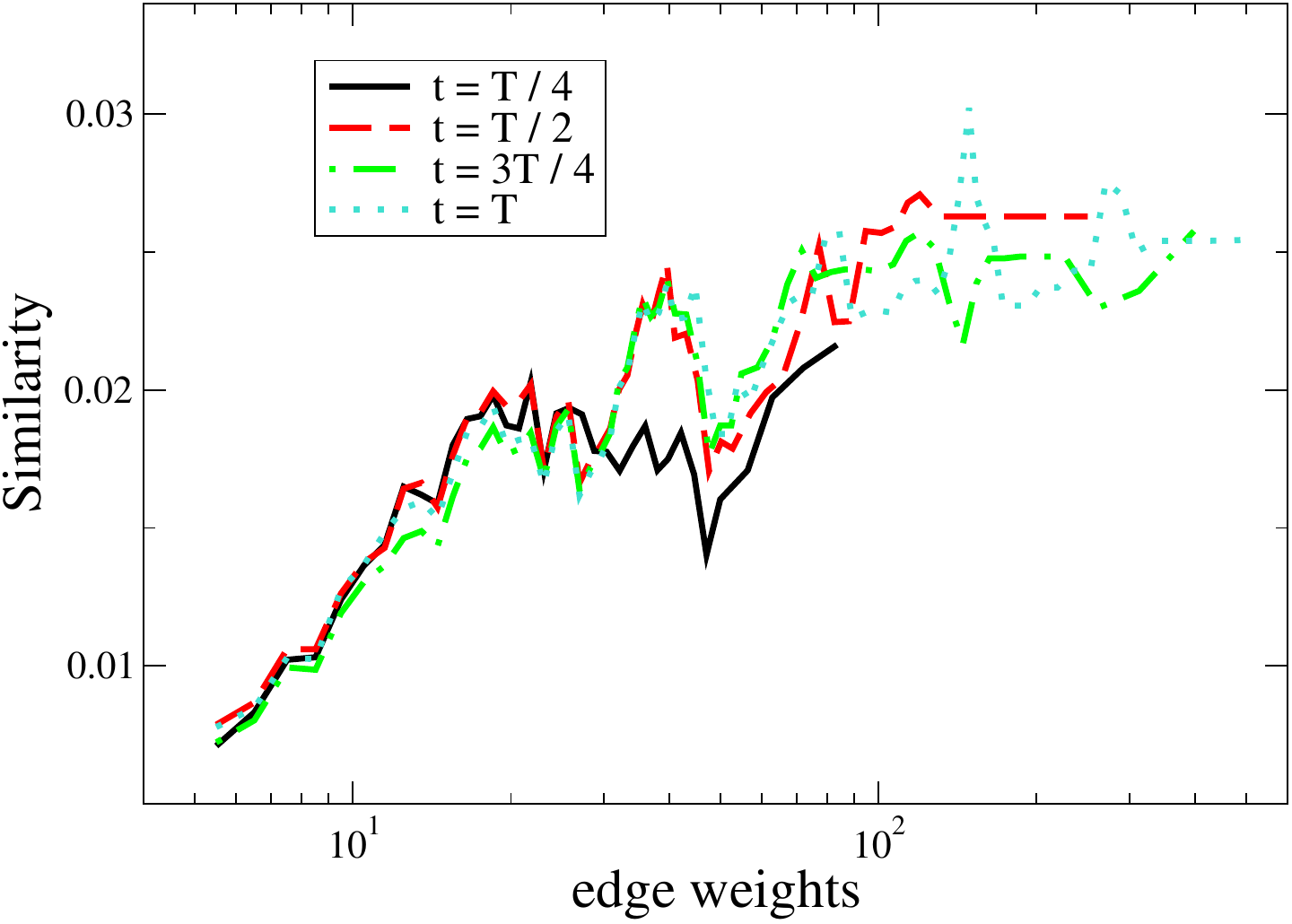}
\caption{\label{fig10}Comparison of mean similarity with edge weights for $9^{th}$ day ($SG_{SECS}$ dataset). The mean similarity of opinions with the edge weights are shown at four different instances of time - $t = T/4$, $t = T/2$, $t= 3T/4$ and $t = T$ where $T$ is the total time. The curves are smoothed by taking sliding window moving average with window size of 10.}
\end{center}
\end{figure}

As an additional objective, in fig~\ref{fig10}, we show how the frequency of interaction between a pair of individuals predicts the similarity of the opinions among the individuals over different instants of time. We measure the similarity of opinions between a pair of individuals by Jaccard Coefficient (JC) of their inventories. It is formally defined as the size of the intersection divided by the size of the union of the inventories i.e., $JC(A_i,A_j) = \frac{|A_i\cap A_j|}{|A_i\cup A_j|}$
where $A_i$ is $i^{th}$ agent's inventory. From all the graphs, it is evident that there is a trend of having higher similarity in opinions between agents showing higher edge-weights where edge-weight reflects the frequency of interactions between an agent pair till that particular instant of time. Thus, with frequent meetings, individuals tend to share similar opinion. This usually also happens in real-life scenarios where more we meet more is the alignment between us on different opinions.

\begin{figure}[t]
 \begin{center}
 \includegraphics*[width=1\columnwidth,angle=0]{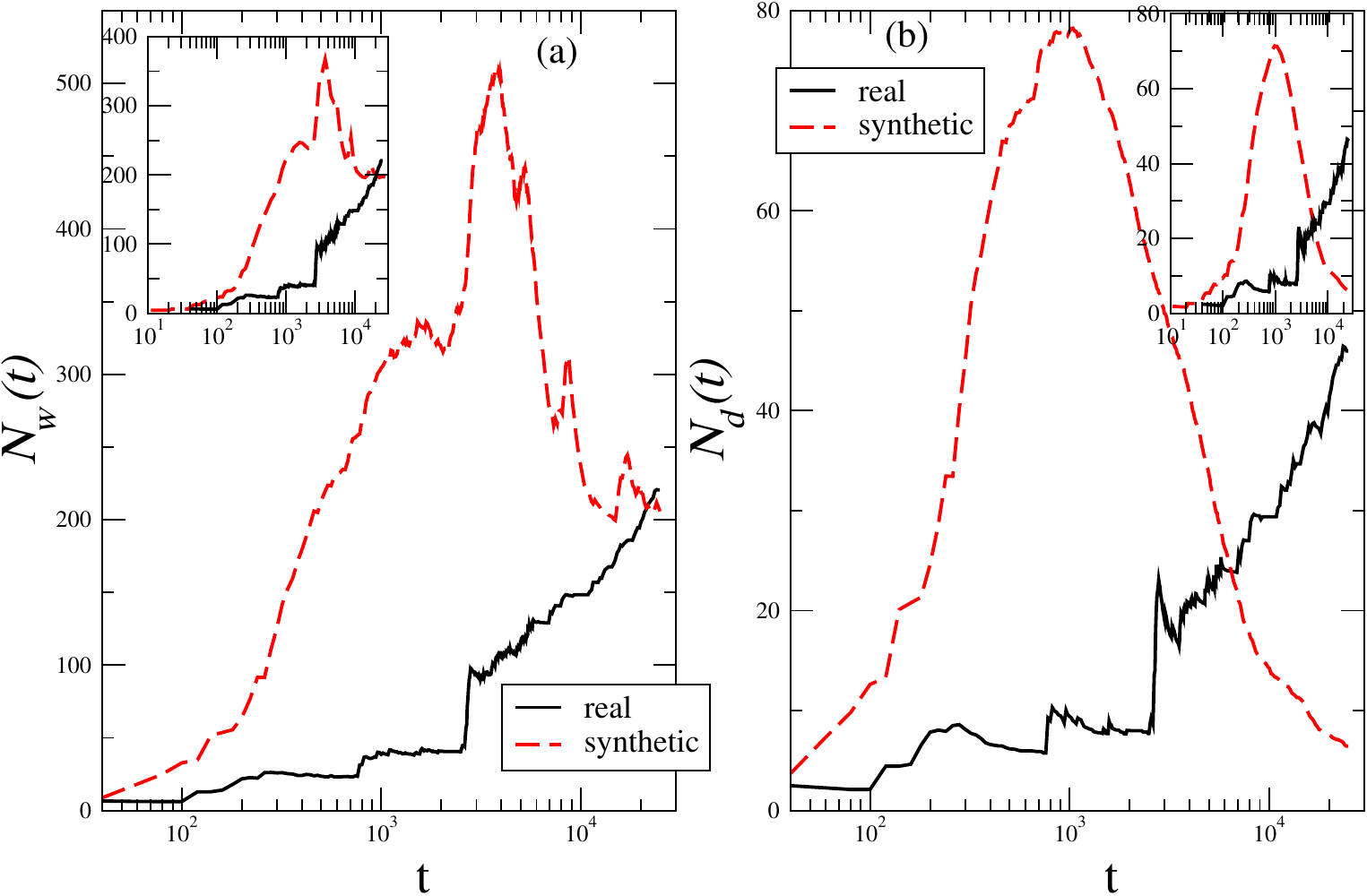} 
\caption{\label{fig11}(Color online) The temporal evolution of (a) $N_w(t)$ and (b) $N_d(t)$ for $SG_{SECS}$ $9^{th}$ day network (both real and synthetic). The datapoints on the curve are averaged over 100 realizations. The number of network realizations for the synthetic case is also 100. The insets show $N_w(t)$ and $N_d(t)$ behavior for control experiment keeping degree sequence same at each time instance.}
 \end{center}
\end{figure}

\textit{Control experiments:} 
For the purpose of control experiment, we create simulated versions of the time-varying snapshots of the $9^{th}$ day network. We randomize the link structure of the empirical real networks at each 20s time interval preserving the average degree of the network. Note that, this method destroys the temporal
correlations of successive contact sequences. There have been already some studies as to how to generate synthetic versions of time-varying networks. Starnini et al.~\cite{rand} have put forward three different approaches for synthetic extension of contact sequences preserving certain statistical properties of the real network. The first approach is sequence replication where the contact sequences are extended by repeating them periodically. The second approach is sequence randomization which randomizes the time ordering of the contact sequences. 
The third approach is statistically extended sequence which generates the sequence by choosing the average number of new connections starting at that time step, randomly from all the conversations. Our method is to some extent similar in spirit to the third approach. We also perform the control experiments preserving the exact degree sequence at each time instant while randomizing the agent ids of the agent who take part in the game at that time instant; nevertheless, the behavior of the observables tend to remain very similar to the previous method (see fig~\ref{fig11} insets). We remark that a detailed analysis of the naming game dynamics on the different variants of synthetic networks is an interesting direction of future research.

\begin{figure}[h]
\begin{center}
 \includegraphics*[width=1\columnwidth,angle=0]{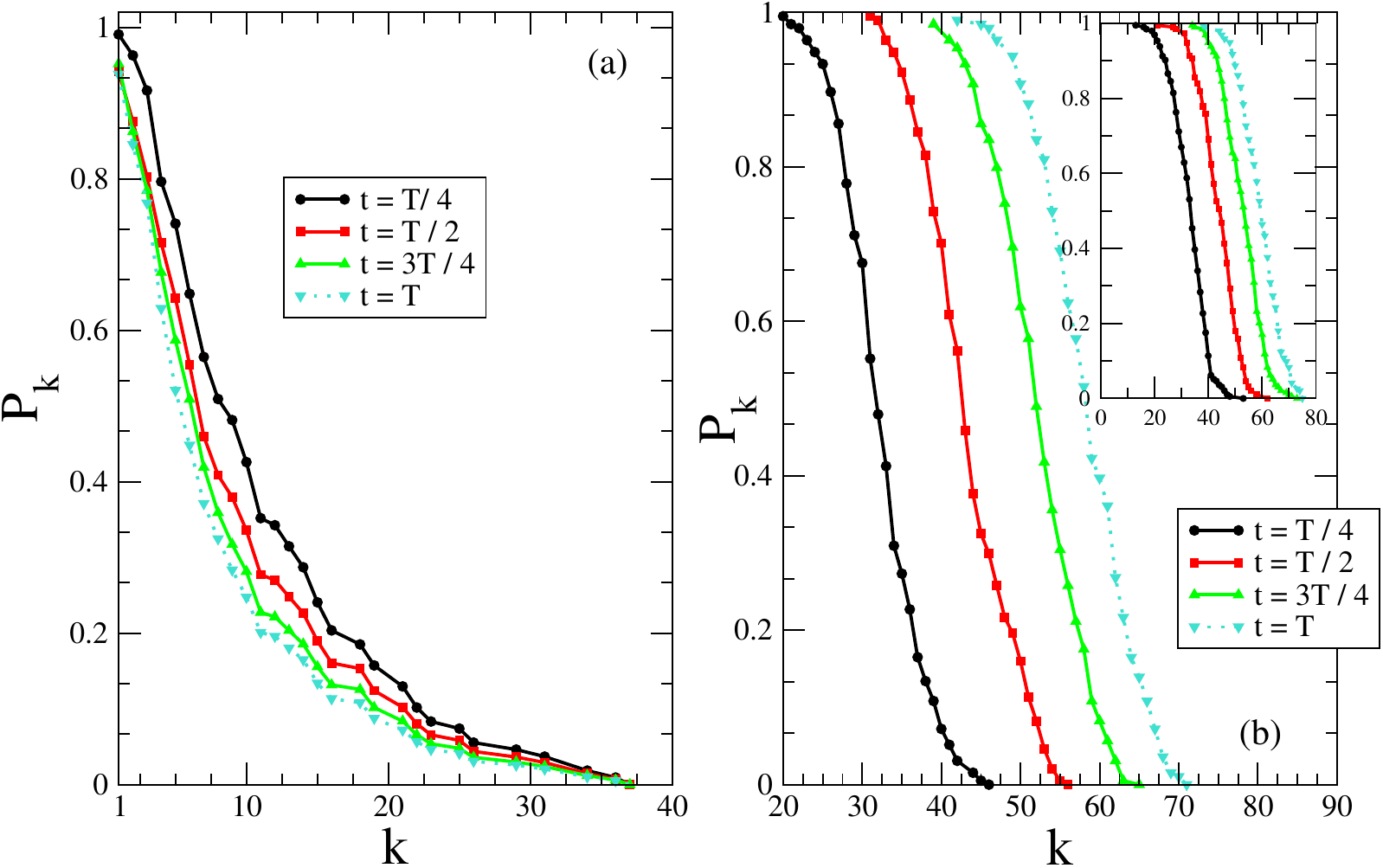}
 \caption{\label{fig11.9}(Color online) (a) Cumulative degree distribution at different time instants: $t = T/4$, $t = T/2$, $t= 3T/4$ and $t = T$ for the real network of day 9 (b) Cumulative degree distribution at different time instants: $t = T/4$, $t = T/2$, $t= 3T/4$ and $t = T$ for the corresponding synthetic networks. The inset shows the degree distribution for control experiment keeping degree sequence same at each time instant.}
\end{center}
\end{figure}
The behavior of the $N_w(t)$ and $N_d(t)$ (see fig~\ref{fig11}) are not in the lines of what we observe in the real counterparts. In all the simulated networks, the $N_w(t)$ and $N_d(t)$ behave similarly as in case of static Erd\H{o}s-R\'{e}nyi graphs~\cite{luca}. This differences in the NG dynamics is due to the fact that the link structure and the time correlation gets completely lost in synthetic network. This is well supported by the differences in degree distribution at different time instances ($t = T/4$, $t = T/2$, $t= 3T/4$ and $t = T$) for the real and synthetic network (see fig~\ref{fig11.9}).

\subsubsection{Results from the $HT_{SECS}$ dataset}
In this section, we consider the second dataset containing dynamic face-to-face interactions among 113 conference attendees. 
We first study the global behavior of the system through the temporal evolution of three main quantities: the total number $N_w(t)$͒ of opinions in the system, the number of different opinions $N_d(t)$͒, and the rate of success $S(t)$. 

\begin{figure}[h]
\begin{center}
 \includegraphics*[width=1\columnwidth,angle=0]{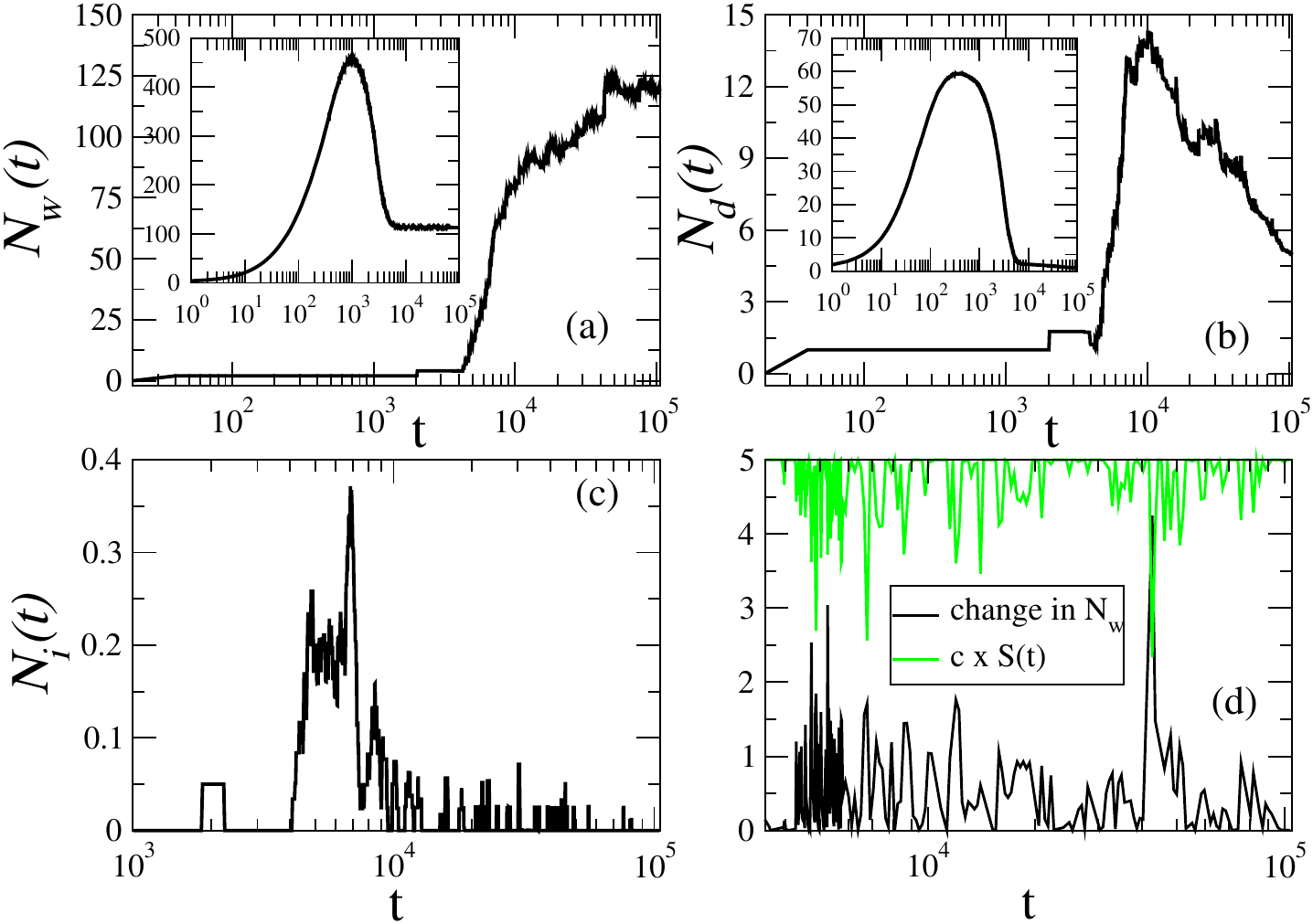}
\caption{\label{fig12}(Color online) The temporal evolution of (a) $N_w(t)$ and (b) $N_d(t)$ on time-varying conference network respectively. The insets show the evolution of $N_w(t)$ and $N_d(t)$ on the static and composite version. The data are averaged over 1000 simulation runs. 
(c) The number of inventions of opinions $N_i(t)$ over time. (d) Comparison of $\Delta N_w(t)$ with success rate $S(t)$. }
\end{center}
\end{figure}
The curve corresponding to $N_w(t)$ shows an initial slow growth followed by a sharp transition and finally reaching a steady growth regime (see fig~\ref{fig12}(a)). Note that this result is markedly in contrast to what would have been observed if the games were played on the composite network
 constructed at the end of the conference (see fig~\ref{fig12}(a) inset). In fact, this result is in contrast to most of the other results that have been reported in the literature so far indicating that the time-varying nature of the underlying societal structure has a strong impact on the emergent pattern of opinion formation.
Similar trends are also observed for $N_d(t)$ - initially a slow growth followed by a sharp transition reaching a peak and finally a steady descent. However, note that the value of $N_d(t)$ does not become one (i.e., convergence criteria is not reached) even after $10^5$ games have been played (see fig~\ref{fig12}(b)). 
This is strikingly in contrast with the case of composite network (see fig~\ref{fig12}(b) inset) where $N_d(t)$ becomes one as early as within $10^4$ games.

Initially, as time proceeds, new individuals join the network that increases the number of inventions of new opinions (see fig~\ref{fig12}(c)) thus causing a rise in both $N_w(t)$ and  $N_d(t)$. However, later on new inventions stop (fig~\ref{fig12}(c)) as the players joining late are less compared to the number that have already 
joined and are therefore rarely chosen as speakers thus inhibiting new inventions. Hence, $N_d(t)$ is found to drop in the later stage of the dynamics although pointing to a clear existence of multiple opinions. In contrast, $N_w(t)$ doesn't drop because although new opinions are not formed, old opinions trapped in different groups 
do not get disposed off the system.

\begin{figure}[h]
\begin{center}
\includegraphics*[scale=0.5]{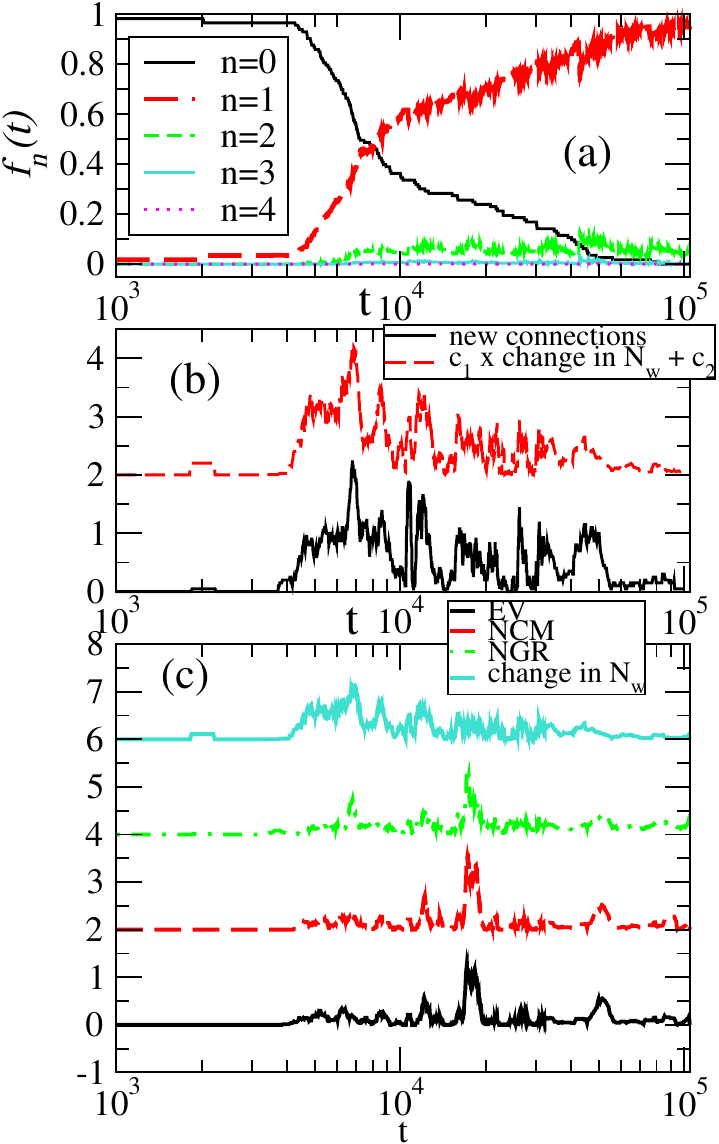}
\caption{\label{fig13}(Color online) (a) Evolution of inventory sizes $n$ $(n = 0, 1 . . . )$. $f_n(t)$ is the fraction of agents whose inventory size is $n$ at time $t$. (b) Comparison of temporal evolution of $\Delta N_w(t)$ and the number of new connections smoothed by taking sliding window moving average with window size 20.
 (c) Comparison of $\Delta N_w(t)$ with the variance of community sizes (found by NGR, NCM and EV algorithm) evolving over time (the curves are suitably scaled by some constant for the purpose of better visualization). The data are smoothed by taking sliding window moving average with window size 20.}
\end{center}
\end{figure}
Further, in fig~\ref{fig12}(d) we show how the absolute change in $N_w(t)$ is driven by the rate of success of agents; $\Delta N_w(t)$ increases with a decrease in $S(t)$ while it decreases with an increase in $S(t)$. Finally, an important analysis that is required to complete the picture centers around the precise reason for the steady growth in 
$N_w(t)$ in the final regime of the dynamics. We attempt to provide a plausible explanation for this through a series of results reported in fig~\ref{fig13}.

 In fig~\ref{fig13}(a), we present the fraction of agents having 0, 1, 2 and more opinions in their inventories. Clearly, with the evolution of system, the fraction of agents with 
inventory size 0 diminishes; fraction of agents with size 1 increase steadily while that with size 2 is roughly stable; even larger size inventories appear only rarely in the course of the evolution. In addition, we observe that $\Delta N_w(t)$ has a direct correspondence with the number of new connections acquired by the network at each time step
(see fig~\ref{fig13}(b)). These new connections trigger an increase in failure events, thereby increasing $N_w(t)$; at the same time success events cannot reduce $N_w(t)$ since in most cases the inventory sizes of the agents are already very low ($\sim 1$) indicating that most of these success events are actually again ``success with no outcome''.
 This last observation points to the fact that there should be an inherent community structure in the time-varying network and this is made apparent through fig~\ref{fig13}(c) where we report the variance of the size of the communities (using three different algorithms as in the previous cases) and show that this is highly correlated to $\Delta N_w(t)$.
 In summary, the presence of community structure coupled with a continuous influx of new connections (leading to late-stage failures in the system) together lead to the steady growth of $N_w(t)$ in its final regime of evolution.

\begin{figure}[t]
\begin{center}
\includegraphics*[scale=0.3,angle=0]{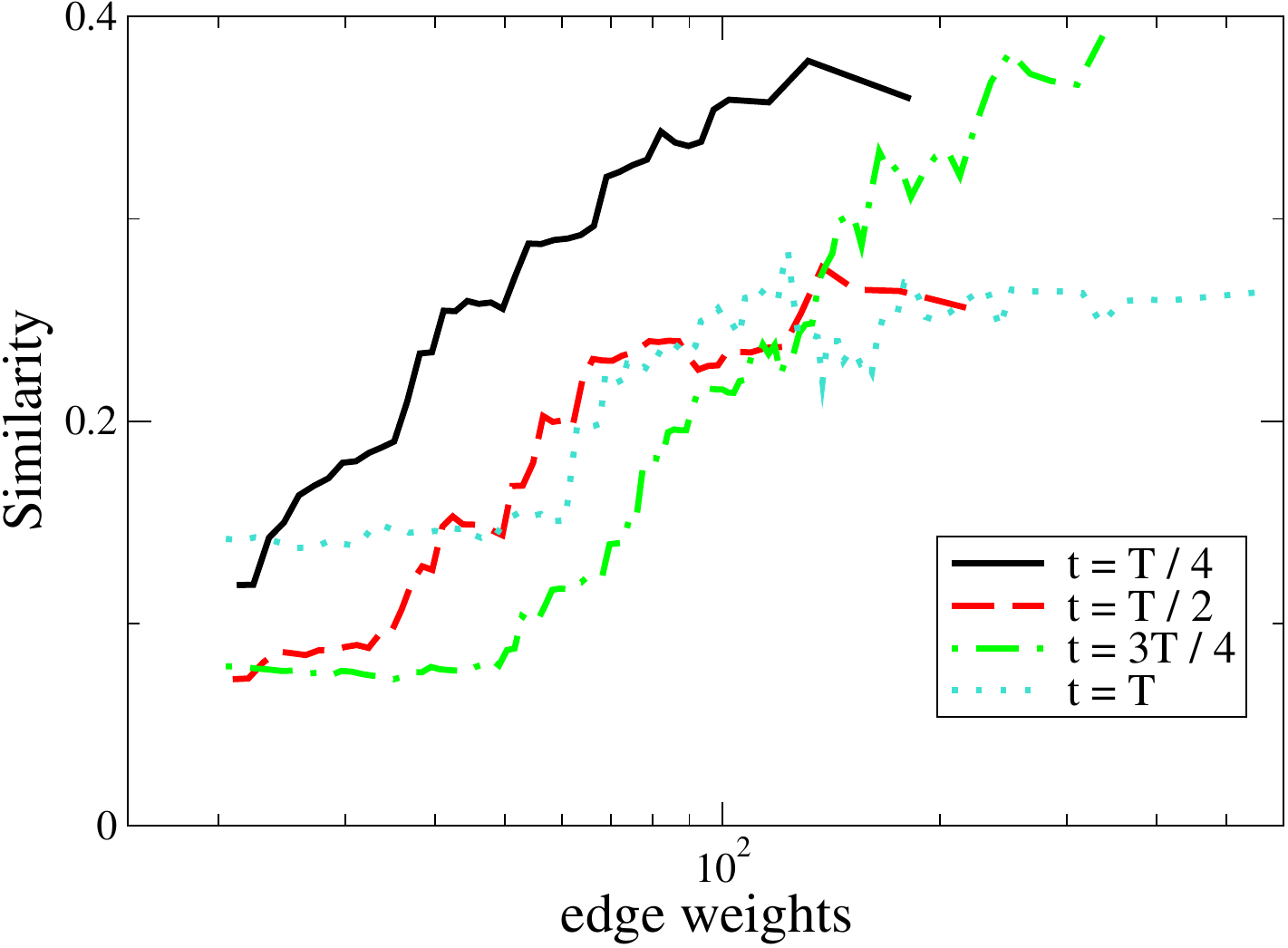}
\caption{\label{fig14}Comparison of mean similarity with edge weights for $HT_{SECS}$ dataset. The mean similarity of opinions with edge weights are shown at different instances of time - $t = T/4$, $t = T/2$, $t= 3T/4$ and $t = T$ where $T$ is the total time.
 The curves are smoothed by taking sliding window moving average with window size 40.}
\end{center}
\end{figure}

In fig~\ref{fig14}, we present mean similarity of opinions among individuals with edge weights at different time instances. In all the instances, there is a positive correlation between similarity of opinions and the frequency of interactions (edge-weights).

\textit{Control experiments}:
For the $HT_{SECS}$ dataset also, we create synthetic networks using the method outlined previously. We study the two most important observables $N_w(t)$ and $N_d(t)$ by playing naming game on these simulated networks. Both these quantities show a different behavior from their real counterpart.
 The $N_w(t)$ and $N_d(t)$ in the simulated networks show two distinct two regions - a steady growth and then a fall whereas the $N_w(t)$ curve in the real network show a slow growth zone followed by a sharp transition and finally a zone of steady growth. The simulated networks tend to behave as standard Erd\H{o}s-R\'{e}nyi graphs. The difference in the behavior can be precisely attributed to the loss of the link structure and time correlations as in the previous case. 
\begin{figure}[h]
 \begin{center}
 \includegraphics*[width=1\columnwidth,angle=0]{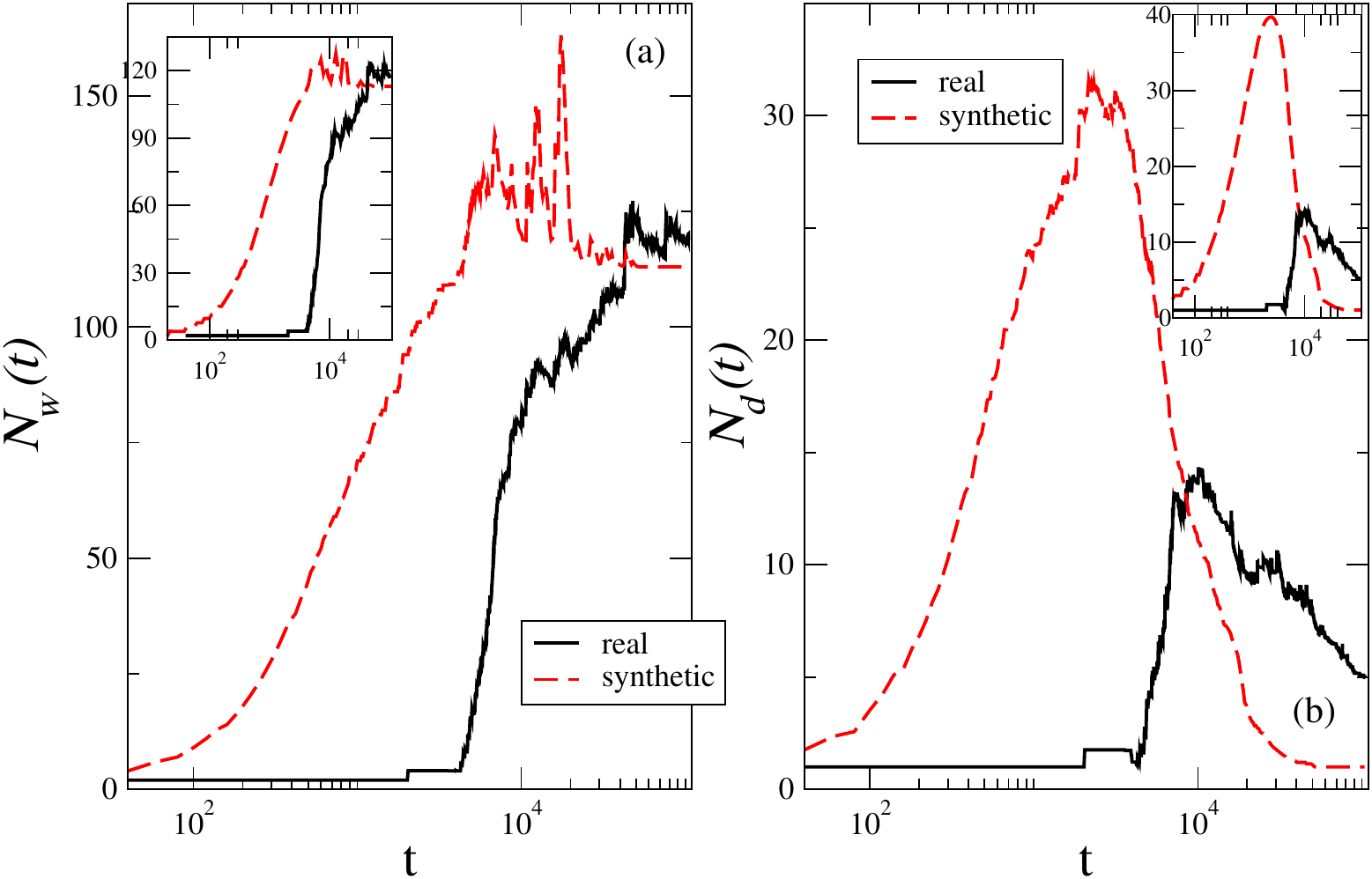} 
\caption{\label{fig15}(Color online) Comparison of the global quantities in the real and the simulated networks. Temporal evolution of (a) $N_w(t)$, (b) $N_d(t)$ for $HT_{SECS}$ real and synthetic networks. The datapoints on the curve are averaged over 100 simulation runs. Number of the network realizations for the synthetic case is also 100. The insets show $N_w(t)$ and $N_d(t)$ behavior for control experiment keeping degree sequence same at each time instant.}
 \end{center}
\end{figure}

\section{Conclusions and future work}
 In this paper, we studied the Naming Game as a model of opinion formation on the time-varying social networks. While considering composite snapshots accumulated over a certain period (e.g., 69 instances of $SG_{DAY}$ dataset) both the maximum memory ($N_w^{max}$) and the time to reach this memory peak ($t_{max}$) scale as $O(N)$; however, the time to reach the consensus strongly depends on the presence of community structure (rather than a straight-forward $N^{1.4}$ scaling).
We also considered the time-evolution of the network in perfect synchronization with the steps of the game (e.g., $SG_{SECS}$ and $HT_{SECS}$ dataset) and found that the emergent behavior of the most important observables (i.e., $N_w(t)$ and $N_d(t)$) have a nature that is markedly in contrast to what has been reported so far in the literature thus indicating the strong influence of the underlying societal structure on the dynamics of opinion formation.
 While in case of $SG_{SECS}$, we observe that new inventions along with a continuous influx of new agents keeps both $N_w(t)$ and $N_d(t)$ sharply growing, in case of $HT_{SECS}$, successful interactions among older agents cause inventions to stop (hence a fall in $N_d(t)$) although late-stage failures continue to exist due to influx of new agents thus contributing to a steady growth of $N_w(t)$ in the final phase of the dynamics. The fall of $N_d(t)$ curve is not observed in case of $SG_{SECS}$ possibly because in this case the games are played over a shorter span of time (1 day)
in comparison to $HT_{SECS}$ where the games are played over 2.5 days so that enough successful interactions could be realized.

There are quite a few interesting directions that can be explored in the future. One such direction could be to incorporate the dominance index of the agents into the model. Not all actors in a society are equally dominant; while some of the actors are more opinionated and dominant the others might be more passive. This characteristic property can be incorporated into the model by ranking those agents that are more successful in their past interactions
as more dominant. In this setting, it would be interesting to investigate the scaling relations most naturally under the constraints that the dominant agents are allowed to speak more. Another direction could be to investigate the effect of the flexibility of the agents in adapting to new opinions (traditionally modeled by a system parameter $\beta$ that encodes the probability with which the agents update their inventories in case of successful interactions~\cite{Baronchelli07})
 when they are embedded on time-varying networks. Finally, a thorough analytical estimate of the important dynamical quantities reported only through empirical evidence here is needed to have a ``clear-cut'' understanding of the emergent behavior of the system. In fact, there has been some progress already made in this direction in the form of the study that focuses on the consensus of mobile agents using naming game dynamics with local broadcasts~\cite{mobileNG}. Our approach differs with this model in the following way: real world networks are 
formed by interactions with agents arriving and leaving at different points in time whereas the mobile agents of~\cite{mobileNG} moving on the two-dimensional plane always remain within the system throughout. The networks that are built from these randomly moving agents are very much different from real world networks which show different topological properties that are well distinguished from random networks. Nevertheless, coupling the concept of mobility of agents in the form suggested in ~\cite{mobileNG} on top of the time-varying structure presented here can lead to novel and interesting insights and
is a part of future investigation.

\bibliographystyle{unsrt}
\bibliography{refpre}

\end{document}